# Solid-State Effects on the Optical Excitation of Push−Pull Molecular J-Aggregates by First-Principles Simulations

Michele Guerrini,[†,‡] Arrigo Calzolari,[‡] and Stefano Corni*,[‡,§]

[†]Dipartimento FIM, Università di Modena e Reggio Emilia, 41125 Modena, Italy
[‡]CNR Nano Istituto Nanoscienze, Centro S3, 41125 Modena, Italy
[§]Dipartimento di Scienze Chimiche, Università di Padova, 35131 Padova, Italy

**Ⓢ** Supporting Information

**ABSTRACT:** J-aggregates are a class of low-dimensional molecular crystals which display enhanced interaction with light. These systems show interesting optical properties as an intense and narrow red-shifted absorption peak (J-band) with respect to the spectrum of the corresponding monomer. The need to theoretically investigate optical excitations in J-aggregates is twofold: a thorough first-principles description is still missing and a renewed interest is rising recently in understanding the nature of the J-band, in particular regarding the collective mechanisms involved in its formation. In this work, we investigate the electronic and optical properties of a J-aggregate molecular crystal made of ordered arrangements of organic push−pull chromophores. By using a time-dependent density functional theory approach, we assess the role of the molecular packing in the enhancement and red shift of the J-band along with the effects of confinement in the optical absorption, when moving from bulk to low-dimensional crystal structures. We simulate the optical absorption of different configurations (i.e., monomer, dimers, a polymer chain, and a monolayer sheet) extracted from the bulk crystal. By analyzing the induced charge density associated with the J-band, we conclude that it is a longitudinal excitation, delocalized along parallel linear chains and that its overall red shift results from competing coupling mechanisms, some giving red shift and others giving blue shift, which derive from both coupling between transition densities and renormalization of the single-particle energy levels.

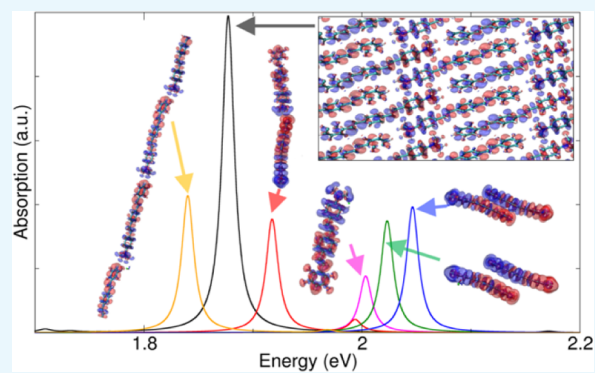

## ■ INTRODUCTION

J-aggregates[1] are a general class of low-dimensional molecular dye crystals, which display coherent interaction with light. These systems have interesting optical properties such as an intense and narrow (i.e., long radiative lifetime) red-shifted absorption peak (called J-band) not present in the spectrum of the single monomer unit they are composed of, that enhances both emissive and nonlinear optical properties.[2] Furthermore, because of the intermolecular mechanisms (i.e., van der Waals, π−π stacking, hydrogen bonding, etc.), they have the ability to delocalize and migrate electronic excitations at long distances.[3,4] The position, intensity, and width of the red-shifted peak is strongly related to the molecular arrangements and their mutual interactions.[4−6]

Since their discovery in 1936 by Jelley,[1] J-aggregates have been widely studied and characterized in the literature, with a number of technical analyses about their dimensions, shapes, and morphologies. From the experimental point of view, they are prepared and formed in solutions (e.g., water) and extended three-dimensional (3D) crystals are not usually observed. Rather, they often assemble in low-dimensional structures. At the mesoscale, J-aggregates manifest a complex morphological and structural variability, so that various shapes have been observed at different dye concentrations and solution conditions.[2,5−9] Moreover, these molecules tend to aggregate in herringbone, brickwork, ladder/staircase structures, with molecular planes normally oriented to the monolayer plane. A detailed discussion about the links between J-aggregates morphologies and optical properties is in ref 10.

In parallel, several theoretical models have been put forward to explain the characteristic narrowing and enhancement of red-shifted band of J-aggregates.[11−17] Because vibronic effects play no relevant role in the J-band,[18] these models are often simplified by assuming the adiabatic Born−Oppenheimer approximation and weak overlap between molecular wave-functions, as well as small excitations.[11] This description has been extensively used to reproduce the excitonic energy transfer in monomolecular chains.[19,20] In other related works,[18,21] the position of the red-shifted peak is predicted from the knowledge of the monomer absorption spectrum and the interaction strength between the monomers. The latter model has been proved useful in particular to estimate the molecular packing arrangements from measured absorption spectra.

Despite the huge amount of experimental data, on the theoretical side some crucial problems remain unsolved: to date, the most accurate simulations have described not really extended systems but rather small clusters,[22] with just a few tens of monomers (∼20), whereas experimental samples are composed of hundreds or thousands of units. Solid-state approaches, which take thoroughly into account long-range and many-body effects, have never been proposed. Thus, a *first-principles* description of extended low-dimensional J-aggregates, such as films and (quasi) one-dimensional (1D) chains, would shed light on the complex interplay between the intermolecular coupling and the resulting optical response. The goal is to provide a microscopic rational of how the various molecular arrangements that coexist in a realistic J-aggregate crystal, as well as the quantum confinement along a given direction, contribute to the overall red shift of the peak. This may also help to identify guidelines for the design of J-aggregate with specific optical properties.

This is particularly relevant in view of the renewed experimental interests about these systems, which exhibit an ultrastrong and coherent light–matter interaction, useful to probe objects at the nanoscale. For this reason, J-aggregates have been recently employed together with plasmonic antennas to study hybrid nanostructures in the strong-coupling regime.[23−27] For example, the coupling between organic J-aggregates with metallic nanosystems (e.g., nanoparticles, nanorods, nanodisks, etc.) forms surface-plasmon exciton hybrid states (plexcitons) that could pave the way toward active plasmonic devices operating at room temperature.[28−33]

In this work, we investigate a J-aggregate molecular crystal composed of 4-(N,N-dimethyl-amino)-4′-(2,3,5,6-tetrafluorostyryl)-stilbene.[34]

This molecule is also a push–pull organic dye (Figure 1): it is a π-conjugated system that possesses an intrinsic electric dipole because of the electron donating (push) amine group and the electron withdrawing (pull) F substituents and manifests an intramolecular charge-transfer (ICT) behavior when optically excited.[35−37] The latter is a key property to polarize the molecular crystal because it permits to transfer energy from one molecule to the other and to delocalize the excitation throughout the whole aggregate. The unusual combination of J-aggregate coupling and push–pull character makes this a unique system with fascinating optical properties. Moreover, the choice of this molecule is also motivated by the availability of the X-ray crystal structure (encompassing several kinds of relative molecular arrangements) and optical spectra characterization,[34] and the computational convenience of its being charge neutral (no counterions to simulate) but still electrostatically not trivial due to its ground state dipole moment.

## RESULTS AND DISCUSSIONS

In this section, we present and discuss the results of first-principles time-dependent density functional theory (TDDFT) simulations of a J-aggregate bulk molecular crystal and of two low-dimensional structures, specifically a linear chain and a monolayer film, extracted from it. In particular, we focused on the J-bands and on the associated induced charge densities, to describe the microscopic physical mechanisms involved in the observed red shift. The bulk aggregate is composed of the push–pull organic dye in Figure 1. Because of the presence of the amino substituent on one side and the fluorine atoms on the other, this dye has an intrinsic dipole that polarizes the frontier orbitals, see, for example, highest occupied molecular orbital and lowest unoccupied molecular orbital shown in Figure 1. This is a well-known property of push–pull molecules that imparts an ICT when optically excited.

This property is important to excite the molecular crystal because it permits to delocalize the excitation throughout the whole aggregate. When moving from the isolated molecule to the bulk aggregate (Figure 2), we observe both an enhancement and a red shift of the J-band, in agreement with the experimental evidence.[34] Quantitatively speaking, the experimental red shift turns out to be somewhat larger than what predicted here (0.3 eV vs 0.1 eV), but within the expected accuracy of the calculation. The absolute position of the TDDFT peak is red-shifted (by 0.7 eV) compared with the experiment, as expected for Perdew–Burke–Ernzerhof (PBE) (more in Methods). We do not observe the narrowing of J-band reported in the experiments for extended aggregates because in our simulations the width of the peak has been externally fixed.

To dissect the origin of the red shift, we have first considered three dimer arrangements: a longitudinal ($D_1$) and two stacked dimers ($D_2$, $D_3$), whose atomic positions have been extracted from the experimental X-ray structure of the bulk crystal (Figure 1). In Figure 2, the peak $D_{11}$ is red-shifted with respect to the monomer, whereas $D_2$ and $D_3$ are blue-shifted. These different behaviors can be understood if we analyze the imaginary part of the induced charge density associated with each peak (from now on with the term induced density we will refer to the imaginary part of the induced charge density). In the frequency domain, the imaginary part of the induced charge density represents the portion of the charge that is dephased by $\pi/2$ with respect to the incident electric field and quantifies the optical absorption.[38,39] Thus, assuming a simple dipolar model of the induced densities, the first absorption peak ($D_{11}$ in Figure 2) of the longitudinal dimer

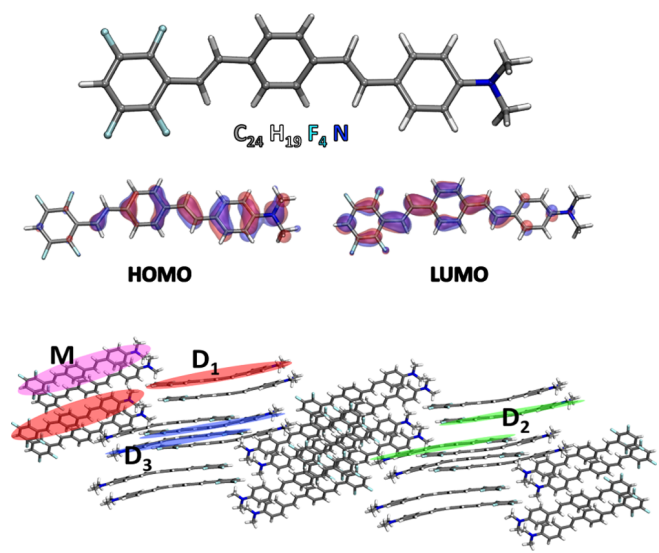

**Figure 1.** (Top) Push–pull dye molecule investigated in this work. Its DFT frontier orbitals are also shown. (Bottom) 3D view of a section of the X-ray structure of the bulk crystal.[34] Here, we have highlighted in different colors the single monomer and the dimer configurations that have been simulated in the present work.

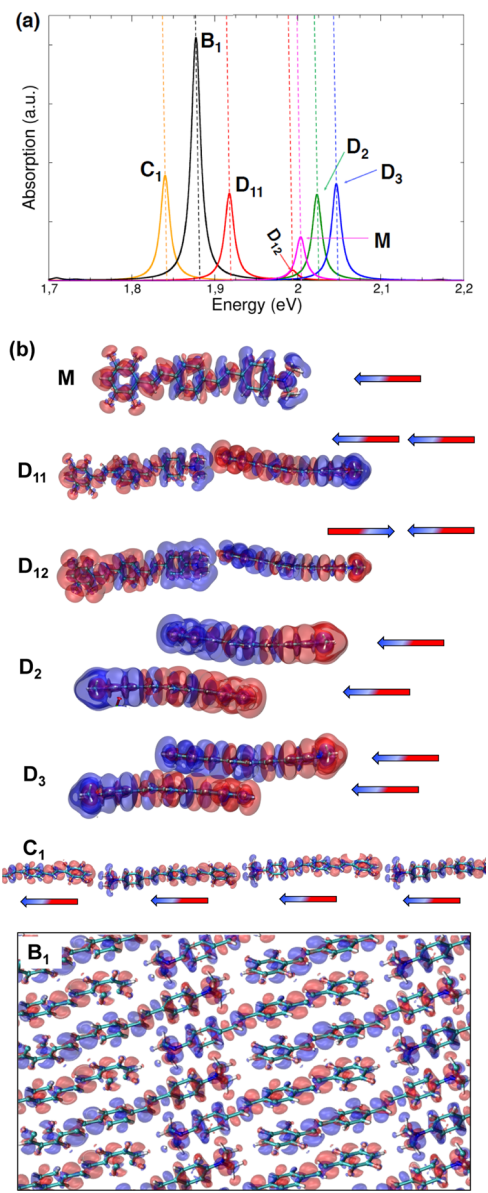

**Figure 2.** (a) TDDFT optical absorption spectra of bulk crystal (black), monomer (magenta), dimers (red, green, blue), and linear chain (ochre) arrangements taken into account. (b) Induced charge density isosurfaces associated with the different absorption peaks selected in the spectra shown in panel (a). The accompanying arrows represent dipole toy models to get a simplified representation of the interaction between the induced charge densities.

($D_1$) is associated with two interacting dipoles in a head−tail configuration.

The origin of the resulting red shift with respect to the isolated monomer is twofold: (i) the coherent Coulomb coupling between the induced charge densities of the two monomers and (ii) the renormalization of single particle energy levels and consequently of the single particle transition energies participating to the total excitation (see Table S.II in section S1 of the Supporting Information). This is due to the ground state Coulomb interaction when monomers are sufficiently close (i.e., in aggregate configuration). In rough terms, the dipole moment of a molecule produces an electric field that acts on those nearby and modifies ground-state properties including single-particle energy levels. These mechanisms contribute to the lowering of the total excitation energy (red shift).

Along the same lines, the induced charge densities of peaks $D_2$ and $D_3$ can be modeled as two stacked and horizontal interacting dipoles. In this case, because of the different orientations (i.e., vertically stacked with a slight lateral shift, see section 2 of the Supporting Information for additional pictorial views) of both induced and static dipoles in each monomer, their interaction causes an overall increase in the total excitation energy. In this configurations, the increase of the total excitation energy can be ascribed to the $\pi-\pi$ interactions. In general, the closer the two dimers are in a vertical stacked arrangement, the higher the blue shift is (see peak $D_5$ in Figure S3 in section 1 of the Supporting Information).

Because the spectra of the isolated dimer configurations extracted from the bulk crystal alone cannot explain the formation of the J-band, we have considered another arrangement that is the periodic version of dimer $D_1$ (red Figure 1): a polymer chain that represents the limit of the bulk crystal if confined in two directions, that is, a 1D J-aggregate. This configuration was revealed to have a J-band even more red-shifted with respect to the bulk. Again, analyzing the overall induced charge density (peak $C_1$ in Figure 2), the red shift can be explained, as before, in terms of intra-chain Coulomb couplings. Because in this case the number of aligned monomers is greater than in $D_1$, both red shift and peak intensity are enhanced. If we consider now the 3D bulk crystal, we observe that the induced charge density of peak $B_1$ in Figure 2 describes an excitation delocalized along parallel chains and with the same crystal symmetry of peak $C_1$ in Figure 2. The corresponding peak is slightly blue-shifted with respect to the linear chain. This is due to the stacking interactions in the bulk (peaks $D_2$, $D_3$) that cause an increase of the excitation energy (i.e., H-aggregate behavior[18,21]).

In addition to the 1D chain, we considered also a two-dimensional (2D) J-aggregate. This 2D film has been obtained from the bulk crystal extracting a monolayer confined along a direction perpendicular to the long axis (see Figure 3 and appendix B of the Supporting Information). Absorption spectra of bulk and monolayer crystals have been reported in Figure 4, whereas the induced charge densities of their J-bands have been plotted in Figure 5.

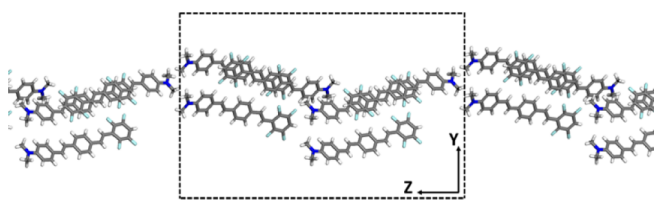

**Figure 3.** Side view of the monolayer film extracted from bulk crystal and confined in one dimension (Y axis) perpendicular to the long axis (Z axis) of the molecular crystal (dashed lines denote the unit cell boundaries).

From 3D bulk to 2D layer, the energy of the J-band changes by a tiny amount. This can be understood by inspecting the corresponding induced charge densities: the confinement of the film does not influence the bulk J-band which is predominantly a longitudinally excitation oriented along the principal crystal axis. Nevertheless, because of the quantum confinement, new peaks at higher energy emerge in the absorption spectrum of the film. In particular, peak $F_3$ in Figure

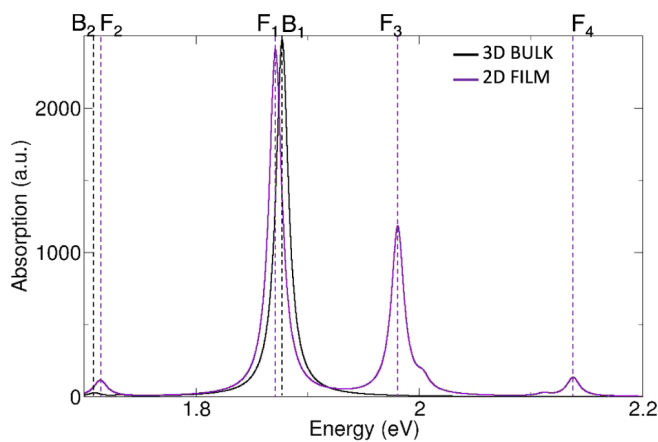

**Figure 4.** TDDFT optical absorption spectra of J-aggregate bulk crystal and monolayer film extracted from it. B1−2 are peaks of bulk crystal, whereas F1−4 are those of the monolayer film.

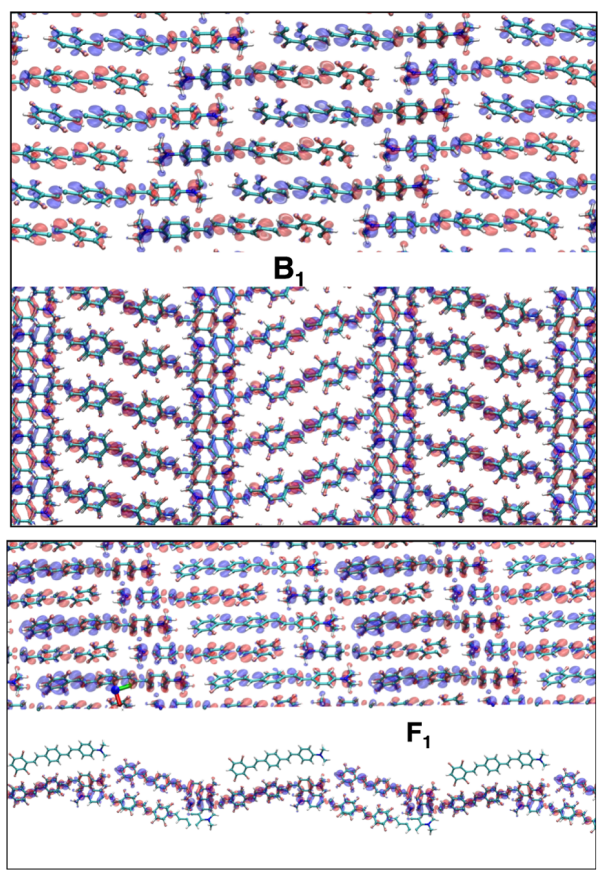

**Figure 5.** Induced charge densities of the main absorption peaks of Figure 4. (Top panel) Induced charge density of peak $B_1$ of bulk crystal; (bottom panel) induced charge density of peak $F_1$ of monolayer film crystal. For each induced charge density, we have plotted two different side views, one is along the plane where both bulk and film are extended and the other along the plane where the film is confined.

4 has an induced density (see Figure S1c in the Supporting Information) whose spatial distribution reflects the broken symmetry of the crystal along the confinement direction. Finally, a small satellite peak ($F_2$) appears at low energy in the spectrum of the film. This peak is present in the spectrum of the bulk crystal as well ($B_2$).

By analyzing the induced charge densities of those peaks (see Figure S1a,b in the Supporting Information), we see that because they do not sum up to zero separately in each monomer (i.e., some of the monomers undergo a depletion, others an increase of charge), these small peaks are expected to have a charge-transfer (CT) character. However, because experiments do not detect any other peak at energies lower than the J-band, we conclude that the simulated CT states are misplaced by the limitations of the adopted semilocal xc functional (PBE), which notoriously underestimates CT excited states.[40]

To conclude, our findings presented in this work allow us to answer most of the questions raised in the introduction:

(1) The behavior of the bulk crystal cannot be deduced considering only its isolated components (i.e., monomer and dimers). Instead, the nature of its optical properties is supramolecular, such that the only way to account for all of the many-molecules related effects is to simulate the extended system.

(2) The microscopic nature of the J-band is analyzed by means of the imaginary part of the induced charge density. The typical red-shifted peak (i.e., J-band) of a J-aggregate can be explained in terms of competitions between stacked and longitudinal molecular arrangements. The coherent Coulomb coupling between the induced charge densities of the longitudinal arrangements (i.e., linear chains) contributes to the red shift, whereas out-of-plane stacking interactions contribute to the blue shift. The latter is weaker than the former, and the overall result is the red shift that characterizes the J-band. In addition, the enhancement of the J-band is discussed in terms of a coherent alignment (along the principal crystal axis) of the transition dipole moments of each monomer in the aggregate, resulting in an overall amplified oscillator strength that gives the enhancement of the intensity of the main peak.

(3) We compared the 3D bulk crystal with two kinds of low dimensional J-aggregate crystals: a monolayer 2D film, confined in one dimension, perpendicular to the principal crystal axis; and a linear 1D chain, confined along two dimensions. These systems allowed us to investigate the major effects of quantum confinement in the extended J-aggregate: in the case of the 2D film, the confinement does not break the symmetry of the J-band, but it generates new excitations at higher energies that, due to surface effects, reflect the broken symmetry of the 3D crystal. In the linear 1D chain case, where the bulk is confined in all but the principal crystal axis, the observed red shift of the J-band is even stronger with respect to that of the bulk. This effect is due to the absence of the stacking interaction that contributes to the blue shift of the main absorption peak.

## ■ METHODS

The geometries of the push−pull molecular dye and unit cell of the bulk J-aggregate (Figure 1) have been taken from the experimental X-ray structure available in the CCDC no. 961738 (for more details see also the Supporting Information of ref 34) with no further relaxations. These are the only external data that have been used.

All of the supramolecular interactions (i.e., van der Waals, π−π, hydrogen bonds, etc.) that contribute to determine the ground-state geometry of the aggregate are implicitly taken into account within the X-ray structure. For the evaluation of the absorption spectra and of the induced charge densities, we have used the Quantum Espresso[41] (QE) suite of codes, based on (time dependent) DFT. In particular, for each system, we have first evaluated the DFT ground state by using the PBE[42] generalized gradient exchange−correlation functional and ultrasoft pseudo-potentials from the PS library.[43] Single-particle wavefunctions (charge density) are expanded in planewaves up to an energy cutoff of 50 Ry (500 Ry).

Only Γ point has been considered for Brillouin zone sampling in the reciprocal space.

The dynamic polarizabilities and the absorption spectra have been simulated by using the turbo-TDDFT code[39,44] of QE, which employs a Liouville−Lanczos superoperator approach for linearized TDDFT.[38] The broadening parameter for each peak has been fixed to 70 meV. The spectra in Figure 2 are normalized following the Thomas−Reiche−Kuhn sum rule in order that the integral over all the frequency range adds up to the total number of electrons per unit cell. For this reason, because in the unit cell of the bulk there are four molecules, its principal peak $B_1$ is about four times that of the monomer (i.e., M) and about two times the peaks of composite molecules (i.e., $D_{11}$−$D_2$−$D_3$) and linear chain (i.e., $C_1$) which contain two molecules per unit cell (see section S2 of the Supporting Information). This explains also why the intensity of peak $C_1$ is about half that of $B_1$. This same reasoning applies to the Figure 4 where the unit cell of the 2D film has a total of six molecules and its absorption spectrum has a larger underlying area with respect to the bulk one.

We exploited the turbo-TDDFT code also for the calculation of the induced charge densities shown in Figure 2b. For each energy, the code returns three different complex-induced charge densities, one for each independent (orthogonal) polarization of the external exciting electric field, see also ref 45 where the same approach has been used. Here, we focus on optical absorption, so we analyze specifically the imaginary parts of the induced densities.

To obtain a unique induced density for each excitation of the system, we evaluated the one that gives the maximum optical absorption. This happens when external electric field and transition dipole moment are maximally coupled and point in the same direction. To achieve that, we have properly linearly combined the three induced charge densities calculated before (see section S3 of the Supporting Information for a clearer explanation).

To simulate the bulk crystal with QE, we have used a monoclinic unit cell containing four monomers with a total of 192 atoms. The isolated structures (monomer and dimers) obtained from the bulk experimental X-ray data were simulated by exploiting periodically repeated supercells, containing the molecular systems in the central position, separated by adjacent replica with ∼10 Å of vacuum in all spatial directions.

For the low-dimensional structures (polymer chain and monolayer film J-aggregates), we applied the same procedure only for the direction(s) along which the system is confined. The monolayer film unit cell has been obtained by extracting a bidimensional layer from the bulk crystal. With the aid of Materials Studio,[46] we have constructed a supercell and then isolated from it a monoclinic unit cell containing six monomer units (288 atoms) and with ∼10 Å of vacuum along the confinement direction (see section S2 of the Supporting Information for crystal structures and computational details).

## ■ ASSOCIATED CONTENT

### *S* Supporting Information

The Supporting Information is available free of charge on the ACS Publications website at DOI: 10.1021/acsomega.8b01457.

Details of absorption peaks of Figures 2 and 4 and the induced charge densities of the extra peaks $B_2$, $F_2$−$F_4$ in Figure 4; additional information about principal transitions involved in the energy peaks of Figure 2 for monomer and dimer configurations and two additional spectra associated with other two dimer configurations extracted from the bulk crystal with the induced charge densities associated with their principal peaks; additional absorption plot referred to the single monomer where we compared PBE and B3LYP xc-functionals; details about crystal structure characterization and computational details of simulated molecular systems; and post-processing procedure we adopted to obtain the induced charge densities associated with each absorption peak (PDF)

## ■ AUTHOR INFORMATION

**Corresponding Author**
*E-mail: stefano.corni@unipd.it.
**ORCID**
Arrigo Calzolari: 0000-0002-0244-7717
Stefano Corni: 0000-0001-6707-108X
**Notes**
The authors declare no competing financial interest.

## ■ ACKNOWLEDGMENTS

The authors thank Prof. Caterina Cocchi of Physics department at Humboldt Universitat Zu-Berlin and Dr. Luca Bursi of Physics and Astronomy department at Rice University Houston (TX) for the useful discussions and the critical reading of the manuscript. This work was partially funded by the European Union under the ERC grant TAME-Plasmons (ERC-CoG-681285).